\documentclass{ws-ijmpcs}

\begin{document}

\markboth{Murphy, Dieckmann, Drury}
{MAGNETIC FIELD AMPLIFICATION \& ELECTRON ACCELERATION IN A PROTOSHOCK}

%
\catchline{}{}{}{}{}
%

\title{FIELD AMPLIFICATION, VORTEX FORMATION, \& ELECTRON ACCELERATION IN A PLASMA PROTOSHOCK: EFFECT OF ASYMMETRIC DENSITY PROFILE
}

\author{GARETH C. MURPHY}

\address{Dublin Institute for Advanced Studies,
Ireland,
gmurphy@cp.dias.ie}

\author{MARK E. DIECKMANN}

\address{ Link\"oping University,
SE-60174 Norrk\"oping, Sweden,
Mark.E.Dieckmann@itn.liu.se}

\author{LUKE O'C DRURY}

\address{Dublin Institute for Advanced Studies,
ld@cp.dias.ie}

\maketitle

\begin{history}
\received{30 Sept 2011}
\revised{Day Month Year}
\end{history}

\begin{abstract}
Gamma ray bursts (GRBs) are thought to originate from highly relativistic jets. The fireball model predicts internal shocks in the jets, causing magnetic field to be amplified \& particles to be accelerated. We model the effects of an asymmetric density configuration for an internal plasma collision in a quasi-parallel magnetic field.
We measured electron acceleration \& found that a tenuous population of electrons is accelerated to Lorentz factors of $\sim$ 300 - close to energy equipartition with ions.
We found that the filaments did not remain static, but were deflected by the Lorentz force \& rolled up into small vortices, which themselves merge to form a larger current vortex.


\end{abstract}

\ccode{PACS numbers: 11.25.Hf, 123.1K}

\section{Introduction}

Prompt emissions of the ultrarelativistic GRBs are probably the most energetic radiative events in the universe. Since the first observations of GRBs, thousands have been detected. They all share a common signature of energetic non-thermal radiation attributed to relativistic motion of electrons in strong magnetic field.
The fireball model predicts internal shocks in the jets, causing magnetic field to be amplified \& particles to be accelerated \cite{Meszaros:1992xq,Spitkovsky:2008rm}. 
The details of the underlying physical mechanism that causes the prompt emissions are still unknown. In particular, how magnetic field is spontaneously generated, \& how electrons are energised to relativistic speeds \& injected into the Fermi mechanism, is under debate. It is therefore relevant to use numerical simulations to probe the behavior of magnetized shocks to see if a robust mechanism to self-consistently amplify the field may be found.
Current models focus on exploiting the symmetry of uniform density plasma cloud collision, in this work, we extend the survey to unequal density plasma clouds.

\begin{figure}[pb]
\includegraphics[width=5cm]{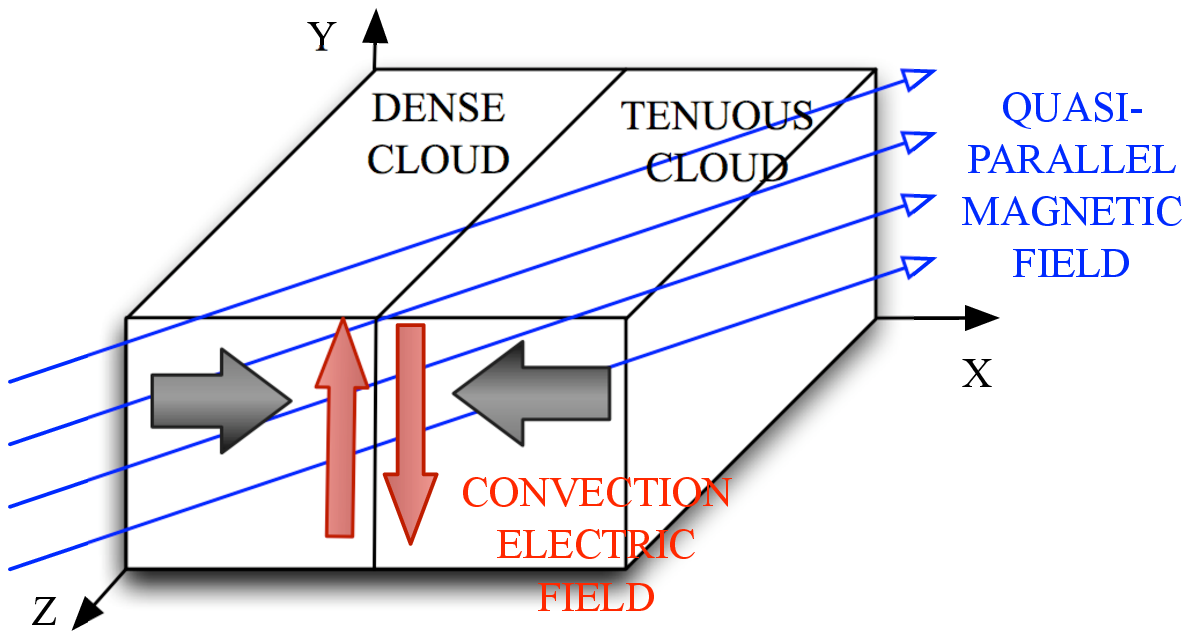}
{\includegraphics[width=7cm]{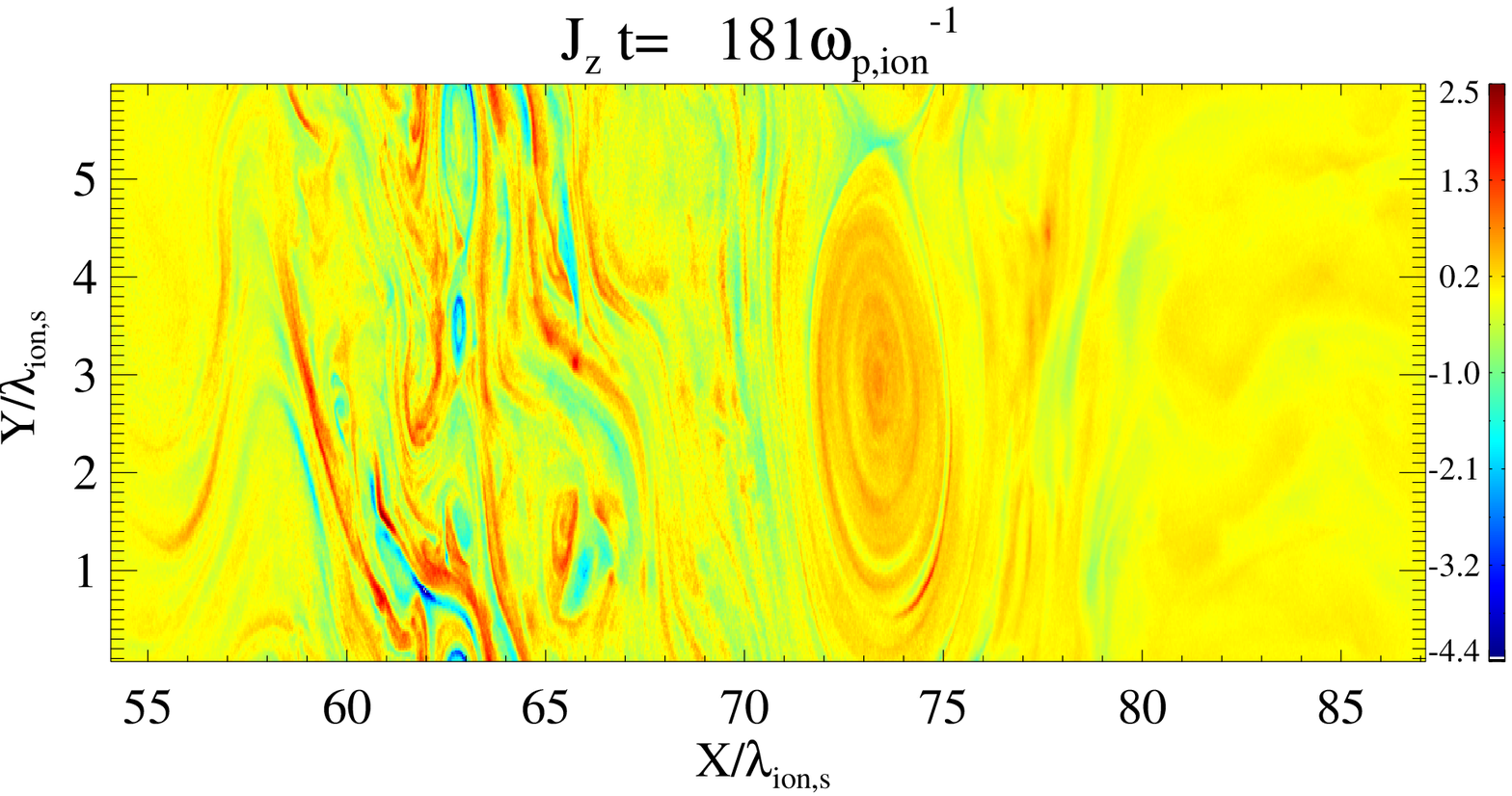}}
\vspace*{8pt}
\caption{Left: Simulation setup. Right: formation of current vortex at collision boundary. \label{f1}}
\end{figure}

{{Two plasma clouds, each consisting of ions and electrons with the mass ratio $m_i / m_e = 250$, collide at 
position $x=0$,   Figure \ref{f1} shows the simulation setup. 
Initial electron and ion number densities 
of the dense cloud both equal $n_1$ and those of the tenuous cloud equal $n_2 = n_1 / 10$. The velocity vectors 
of both clouds are antiparallel and aligned with $x$. The modulus $v_b$ of each cloud in the box frame gives the 
collision speed $v_c = 2v_b / (1+v_b^2/c^2) = 0.9c$. The dense cloud propagates to increasing values of $x$. 
 All species have an initial temperature of 131 keV.
 The magnetic energy to baryonic energy ratio, $\sigma$, is 0.2 (0.02) in the tenuous (dense) cloud. 
The modulus of the convection electric field is $|E_{0y}| = v_b B_{0z}/c$. Both $\nabla \cdot B =0 $ and $\nabla \cdot 
E =0$ at $t=0$. The simulation box size $L_x \times L_y = 656 \lambda_i \times 6 \lambda_i$ is resolved by $2.8 \cdot 
10^4 \times 256$ square grid cells. The dense (tenuous) plasma is resolved by 100 (50) particles per cell per species. 
}}
 
\begin{figure}[pb]
\centerline{\includegraphics[width=0.87\linewidth]{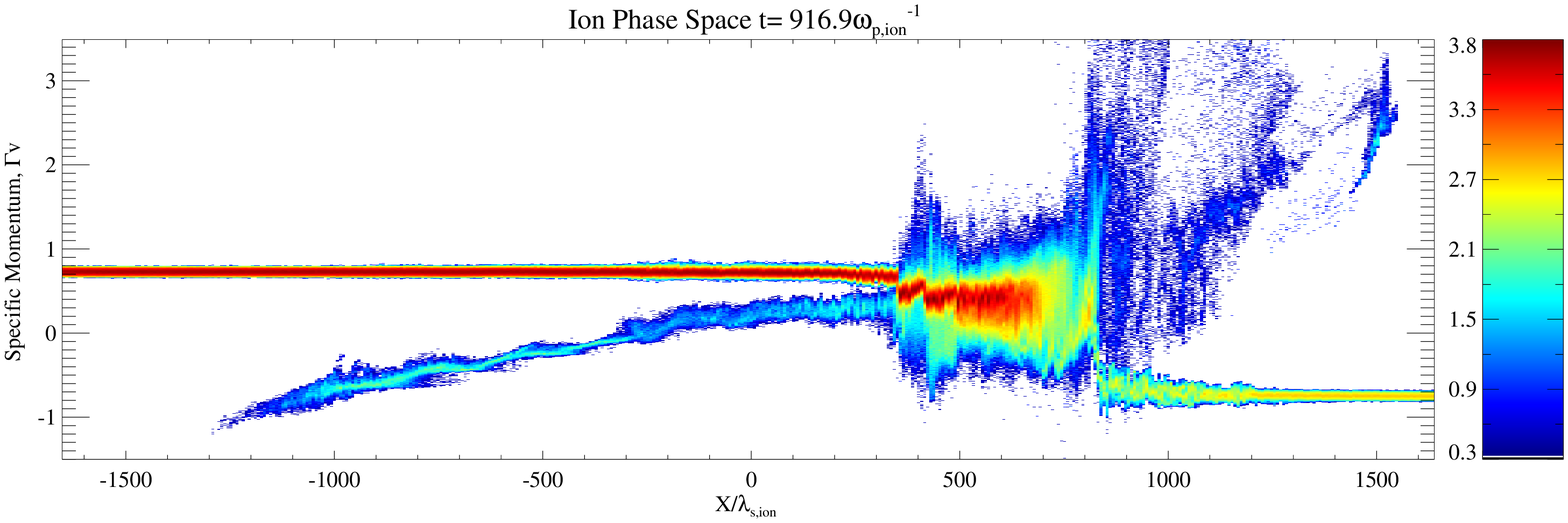}}
\centerline{\includegraphics[width=0.87\linewidth]{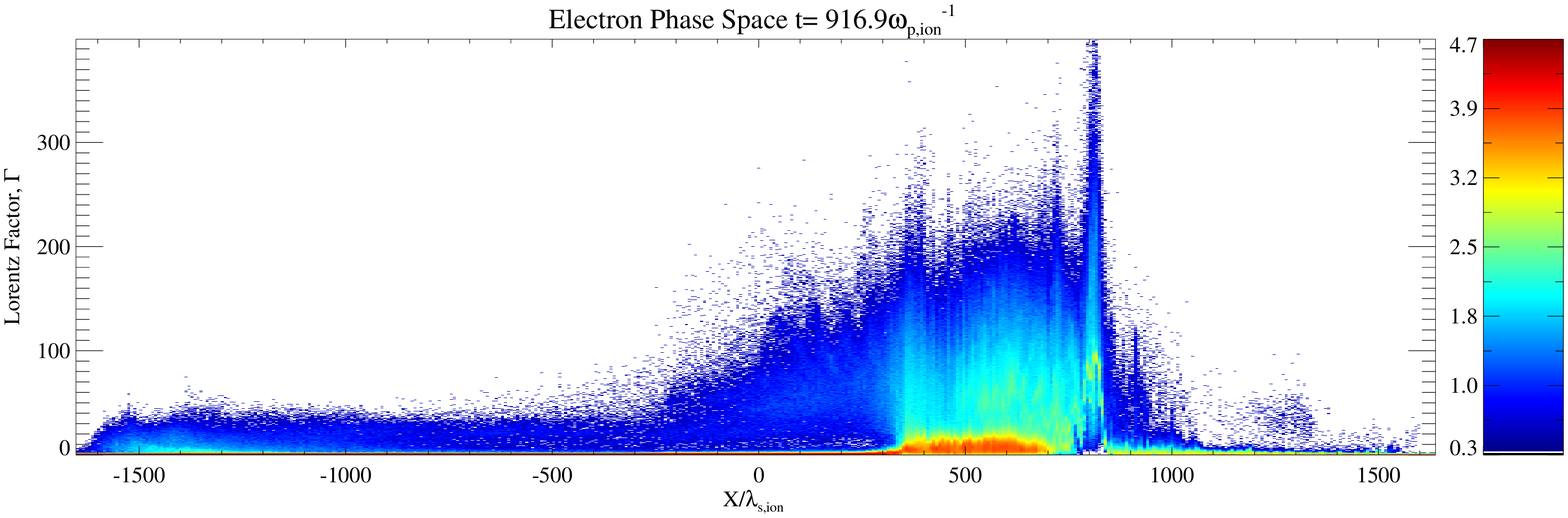}}
\vspace*{8pt}
\caption{Upper panel: Specific x-momentum vs displacement phase space for ion clouds 
Lower panel: Lorentz factor vs displacement phase space for electron species
\label{f2}}
\end{figure}

\section{Results}

{\bf{Vortex formation in plasma shocks:}}
We carried out a large scale, long term 2d relativistic PIC simulation of a plasma collision. 
The peak Lorentz 
factor of the electrons is determined, along with the orientation \& the strength of the 
magnetic field at the cloud collision boundary. The magnetic field component orthogonal to the 
initial plasma flow direction is amplified to values that exceed those expected from the shock 
compression by over an order of magnitude. The forming shock is quasi-perpendicular due to 
this amplification, caused by a current sheet which develops in response to the differing 
deflection of the upstream electrons \& ions incident on the magnetised shock transition 
layer. {{As the upstream plasma impacts on the strong magnetic field,  electrons are deflected away from their original flow direction, while the ion reaction is much weaker. 
Current flows in the (y,z) plane, which amplifies magnetic field. 
The electrons fall behind the ions, because their velocity along x is reduced. An $E_x > 0$ builds up, which
tries to restore the quasi-neutrality. 
Incoming upstream electrons are dragged by it across the magnetic field and accelerated to relativistic speeds.
The distribution of electrons has a non thermal tail with a power law $\sim 2$.}}
 A magnetic 
field structure resembling the cross section of a flux tube grows self-consistently in the current 
sheet of the shock transition layer (Fig. \ref{f1}). Plasma filamentation develops behind the shock front, as
well as signatures of orthogonal magnetic field striping, indicative of the filamentation 
instability. These magnetic fields convect away from  the shock boundary \& their energy 
density exceeds by far the thermal pressure of the plasma. Localized magnetic bubbles form. 
Energy equipartition between the ion, electron \& magnetic energy is obtained at the shock 
transition layer. The electronic radiation can provide a seed photon population that can be 
energized by secondary processes (e.g. inverse Compton).  
We measured electron acceleration \& found that a tenuous population of electrons is accelerated to Lorentz factors of $\sim$ 200 - close to energy equipartition with ions.

We found that the filaments did not remain static, but were deflected by the Lorentz force \& rolled up into small vortices, which themselves merge to form a larger current vortex. In order to check the validity of the 2d approximation we also carried out a short-term simulation in 3d \& found evidence of filamentation. 

{\bf{Long term 1D simulations:}} We carried out long term simulations in 1D in order to verify that the acceleration \& field amplification process was robust \& predict the SSC emission. In 1D, the simulation timescale was increased to 919 $\omega_p^{-1}$. In the 1d simulation, a reverse shock forms, which was not achieved in  2D due to the shorter  timescales. The reverse shock is visible at $x=350$ in Fig. \ref{f1}. Two smaller discontinuouties are seen in the ion phase space plot at $x=400,500$.

\begin{figure}[pb]
\centerline{\includegraphics[width=0.29\linewidth]{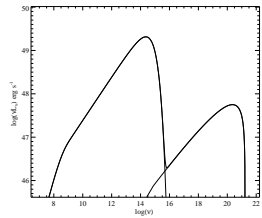}}
\vspace*{8pt}
\caption{Predicted synchrotron self-Compton emission, assuming Doppler factor of 450, magnetic field of 20 nT, maximum Lorentz factor of 1500 using the Compton sphere suite.\label{ssc}}
\end{figure}

{\bf{Synthetic observations:}} In order to analyze the data, a model of the
emission from the hot spot was created using the Compton Sphere Suite \cite{Georg2007}. 
In this one-zone steady-state model of secondary emission processes, a sphere of a given radius is
 permeated by a magnetic field. 
 Photons are produced relativistic electrons via the synchrotron process \& are inverse-Compton scattered by the same electron population to  $\gamma$-ray energies. 
The model assumes synchrotron losses dominate over SSC losses. 
The Thomson scattering cross section is used.
The values from the numerical simulation are used, but must be scaled up from the reduced mass ratio to the true value.
In Fig. \ref{ssc} we plot the derived spectrum.
For future work we shall include the external photon field inverse Compton losses.

{\bf{Conclusions:}}  In 2D simulations the F.I. grows despite the high temperatures \& quasi-parallel background magnetic field (which tends to suppress transverse motion).
  The quasi-parallel field is rotated into a perpendicular field locally at the forward shock transition layer.
  Longer term 1D simulations show similar magnetic field amplification and electron acceleration, while also evolving for sufficient time to see a reverse shock forming.
  The synthetic observations show emission in the MeV range.
  
\section*{Acknowledgments}
GM,LD funded by SFI RFP/ 08/PHY 1694.
H. Ruhl for use of Plasma Simulation Code (PSC).
ICHEC for computer facilities \& support.
GM acknowledges HPC resources (Tier-0) provided by PRACE on Jugene based in Germany.



\begin{thebibliography}{0}    

  \bibitem[1]   {Meszaros:1992xq}
{Meszaros}, P. \& {Rees}, M.~J. 1992, MNRAS, 257, P29

\bibitem[2]{Spitkovsky:2008rm}
{Spitkovsky}, A. 2008, ApJ, 682, L5

\bibitem[3]{2010arXiv1003.1275M}
{Murphy}, G.~C., {Dieckmann}, M.~E.,  Bret, A.,\& {Drury}, L.O'C.  2010{{a}},
A\&A, 524, A84
        
        \bibitem[4]{2010IJMPD..19..707M}
{Murphy}, G.~C., {Dieckmann}, M.~E., \& {Drury}, L.O'C.  2010{{b}},
  IEEE Trans Plasma Sci., 38, 2985


  \bibitem[5] {Georg2007}
  Georganopoulos, M., Kazanas, D., Perlman, E., Wingert,B.,
Graff, P. \& Castro, R. The Compton Sphere,  2007. http://jca.umbc.edu/ markos/cs/index.html

\end{thebibliography}
\end{document}